\documentclass{article}

\usepackage{arxiv}
\usepackage{amsmath}
\usepackage{amssymb}
\usepackage{bbm}
\usepackage{epsfig}
\usepackage{graphics,amsopn,amstext,color,enumerate,bm}
\usepackage{color,float,epstopdf}
\usepackage{mathtools}
\usepackage{svg}
\usepackage{multirow}
\usepackage{graphicx}
\usepackage{url}
\usepackage{microtype}
\usepackage{subfigure}
\usepackage{booktabs} 
\usepackage{hyperref}
\usepackage{wrapfig}
\usepackage{adjustbox}
\usepackage[normalem]{ulem}
\useunder{\uline}{\ul}{}

\newcommand{\FA}{\widetilde{F}}
\usepackage{amsmath, amssymb, amsthm}
\newtheorem{theorem}{Theorem}[section]
\newtheorem{lemma}[theorem]{Lemma}

\usepackage{algpseudocode, algorithm}
\usepackage[noend]{algcompatible}
\usepackage{graphicx}
\usepackage{color}

\DeclareMathOperator*{\argmax}{argmax}

\begin{document}

\title{An Asymptotic CVaR Measure of Risk\\for Markov Chains}

\author{
  Shivam Patel
  \\\texttt{shivamapatel2002@gmail.com} 
  \And 
  Vivek Borkar\\
  \texttt{borkar.vs@gmail.com}
  }
\maketitle

\begin{abstract}
     Risk sensitive decision making finds important applications in current day use cases. Existing risk measures consider a single or finite collection of random variables, which do not account for the asymptotic behaviour of underlying systems. Conditional Value at Risk (CVaR) is the most commonly used risk measure, and has been extensively utilized for modelling rare events in finite horizon scenarios. Naive extension of existing risk criteria to asymptotic regimes faces fundamental challenges, where basic assumptions of existing risk measures fail. We present a complete simulation based approach for sequentially computing Asymptotic CVaR (ACVaR), a risk measure we define on limiting empirical averages of markovian rewards. Large deviations theory, density estimation, and two-time scale stochastic approximation are utilized to define a ‘tilted’ probability kernel on the underlying state space to facilitate ACVaR simulation. Our algorithm enjoys theoretical guarantees, and we numerically evaluate its performance over a variety of test cases.
\end{abstract}

\section{Introduction}\label{introduction}

Conditional Value at Risk (CVaR) has become a standard measure of risk, originally in finance but by now in many other fora, due to its sound axiomatic basis \cite{Artz}. Consider, e.g., a real random variable $X$ with a strictly positive density, so that its distribution $F(\cdot)$ is strictly increasing and continuous. Then CVaR$(\alpha)$ for $\alpha \in (0,1)$ is defined as $E\left[X|X\geq F^{-1}(\alpha)\right]$. (This definition needs some tweaking without the simplifying assumptions regarding the density.) See also \cite{Acerbi}, \cite{Rocka} for equivalent expressions that facilitate convenient computational schemes. Here we consider the problem of evaluating CVaR for a Markov chain over a very long time interval. To ameliorate the computational difficulty thereof, we propose an asymptotic version based on existing results from conditioning with respect to large deviations. This situation has been analyzed and a computational scheme for it has been provided in \cite{Kherani}.
We adapt the results therein to the problem of evaluating Asymptotic CVaR (ACVaR), a concept defined here, and provide theoretical justification for it as well as supporting numerical experiments for the proposed computational scheme. The key step here is to propose a judicious choice of the threshold in \cite{Kherani} that defines the rare event that is being conditioned upon, so as to fit the framework of \cite{Kherani} while maintaining a close relationship with CVaR. This also leads to an additional density estimation step. Our main contribution thus is to formulate a surrogate for the exact CVaR that is computationally reasonable and satisfies at least some of the axioms for a coherent risk measure, and then adapt the scheme of \cite{Kherani} with appropriate modifications for evaluating the same. This is the first step towards a resolution of this problem and points to multiple further research directions that we list in our concluding discussion. For computational scheme for the classical (i.e., non-asymptotic) set-up, see, e.g., \cite{Chow}, \cite{Prash}, \cite{Stanko}, \cite{Mannor}.

We introduce the problem and the key results from \cite{Kherani} in the next section, Section \ref{sec:rare-event-simulation}. Section \ref{sec:acvar} defines Asymptotic CVaR and adapts the framework of \cite{Kherani} to propose a computational scheme for its computation. Section \ref{sec:experiments} presents numerical experiments. Section \ref{sec:conclusion} concludes with a brief discussion.

\section{Markov Chain Simulation Conditioned on Rare Events}\label{sec:rare-event-simulation}

Consider a Markov Chain $\{X_n, n\geq0\}$ on a finite state space $\mathcal{S}$, $|\mathcal{S}| = s$, with an associated reward function  $g(\cdot):\mathcal{S}\to \mathbb{R}$. We are interested in a computable surrogate for large time CVaR defined by
$$E\left[\frac{1}{n}\sum_{m=0}^{n-1}g(X_m)\Big| \frac{1}{n}\sum_{m=0}^{n-1}g(X_m) \geq G_n^{-1}(\alpha)\right],$$
where $G_n$ is the distribution function of $\frac{1}{n} \sum_{i=0}^{n-1} g(X_i)$. The above is simply the classical definition of CVaR specialized to this case, but some non-trivial difficulties with it are already apparent. One is the computational overhead for computing it for increasing $n$. An even more difficult issue is the fact that $G_n$ approaches the  unit step function with a jump at the stationary expectation of $\{g(X_n)\}$ which makes this definition impractical. This suggests looking for suitable asymptotics as $n\to\infty$.  Therefore we consider the limiting quantity 
$$\lim_{n\to\infty}E\left[ \frac{1}{n}\sum_{i=0}^{n-1} g(X_i) \Big|\frac{1}{n}\sum_{i=0}^{n-1} g(X_i) \geq F^{-1}(\alpha)\right]$$
for a suitable surrogate $F$ of $G_n$ for large $n$. Specifically, we choose this $F$ to be the stationary distribution of $g(X_n)$.  The notion of $\alpha^{th}$ percentile conditioning in CVaR does not straightforwardly apply to ACVaR formulation, as the single state cost/reward inverse distribution is a conditioning threshold on a different variable, namely the empirical average of costs/rewards. Intuitively, ACVaR is a relaxation of hard thresholding on individual random variables to a long time average behaviour of the underlying state space. Another simplification we do is to replace the above by
\begin{equation}
\lim_{m\to\infty}\lim_{n\to\infty}E\left[ \frac{1}{m} \sum_{i=0}^{m-1} g(X_i)\Big|\frac{1}{n}\sum_{i=0}^{n-1} g(X_i) \geq F^{-1}(\alpha)\right]. \label{CVAR1}
\end{equation}
This puts it firmly within the framework of the results of \cite{Kherani}. The key result of \cite{Kherani} that broadly falls in the domain of `conditioning on large deviations' (see, e.g., \cite{sadhu1}, \cite{Touchette}) is that \eqref{CVAR1} equals the stationary expectation of $g(\tilde{X}_n)$ for a Markov chain $\{\tilde{X}_n, n \geq 0\}$ governed by a modified transition probability $p^*(i,j), i,j\in \mathcal{S}$, defined by
\begin{equation*}
    p^*(i,j) = \lim_{n\to\infty} P\Big(X_1=j\Big| X_0=i,  \frac{1}{n} \sum_{i=0}^{n-1} g(X_i) \geq \alpha\Big).
\end{equation*}
For this purpose, consider the `multiplicative Poisson equation' that arises in risk-sensitive control, given by
\begin{equation*}
    V_\zeta(i) = \frac{e^{\zeta g(i)}}{\rho_\zeta} \sum_j p(i,j) V_\zeta(j) \hspace{1em} i,j\in \mathcal{S}.
\end{equation*}
where $\zeta>0$. The following results are from \cite{Kherani}.

\begin{lemma} For $\zeta>0$, there exist $V_\zeta>0,\;\rho_\zeta>0$ (with $V_\zeta(\cdot)$ unique up to a constant scaling factor) such that they satisfy the multiplicative Poisson equation, and 
\begin{equation*}
    \ln \rho_\zeta = \lim_{n\to\infty} \frac{1}{n} \ln E\left[exp\left(\zeta \sum_{m=0}^{n-1} g(X_m)\right)\right] \triangleq \Lambda(\zeta).
\end{equation*}
 The map $\zeta \rightarrow \rho_\zeta$ is convex and there exists  for each $\alpha>0$ a unique $\zeta^* = \argmax_{\zeta\geq 0 } (\zeta \alpha - \ln \rho_\zeta)$.
\end{lemma}
\smallskip

These lead to the main result (Theorem 2 and Corollary 2 from \cite{Kherani}) stated next. Let $V^* := V_{\zeta^*}, \rho^* := \rho_{\zeta^*}$. 

\begin{theorem}
The regular conditioned law of $\{ X_m,\;m>0\}$ conditioned on the event $\{ X_0 = x,\; \frac{1}{n} \sum_{k=0}^{n-1} g(X_k)\geq \alpha \}$ converges as $n\to\infty$ to the law of a Markov chain $\{X^*_n\}$ starting at x with transition probabilities $p^*(\cdot,\cdot)$ given by
$$p^*(i,j) := \frac{e^{\zeta^*g(i)}p(i,j)V^*(j)}{\rho^*V^*(i)}, \ i,j\in \mathcal{S}.$$ 
Furthermore,
for any $h:\mathcal{S}\to \mathbb{R}$,
\begin{equation*}
    \lim_{m\to\infty} \lim_{n\to\infty} E\left[\frac{1}{m} \sum_{k=0}^{m-1} h(X_k) \Big| \frac{1}{n} \sum_{k=0}^{n-1} g(X_k) \geq \alpha \right] = E^*[h(X_t^*)]
\end{equation*}
where $E^*[\cdots] :=$ the stationary expectation under $p^*(\cdot,\cdot)$.
\end{theorem}

The proof uses an extension to Markov processes of the Bahadur-Rao theorem for exact asymptotics in large deviations from \cite{Konto}.

\section{Asymptotic CVaR and its Estimation}\label{sec:acvar}

We specialize the foregoing to the case when $h\equiv g$ and motivated by the theoretical results above, define an asymptotic variant of CVaR. Let $F(\cdot)$ denote the limiting cumulative distribution function (CDF) of the reward $g(X_n)$. Since $\mathcal{S}$ is finite, this will be a piecewise constant and non-decreasing function. \\

\textbf{Definition} For any function $g(\cdot):\mathcal{S}\to \mathbb{R}$ and $c\in\mathbb{R}$, the Asymptotic CVaR (ACVaR) is defined as
\begin{eqnarray}
    &&\lim_{m\to\infty} \lim_{n\to\infty} E_x\left[\frac{1}{m} \sum_{k=0}^{m-1} g(X_k) \Big| \frac{1}{n} \sum_{k=0}^{n-1} g(X_k) \geq F^{-1}(c) \right]  \nonumber \\
    &&= \  E_x^*[g(X^*_i)].  \label{acvar}
\end{eqnarray}

\smallskip

In other words, we have set $\alpha = F^{-1}(c)$.  For $X_n \equiv$ a fixed random variable, this would reduce ACVaR to CVaR as desired. In practice,  we can use a suitable estimate of $F^{-1}(c)$ as a surrogate thereof. In the present work, we have used a Gaussian kernel density estimate \cite{Chen}. The actual stationary distribution of the reward, being a function of a finite state Markov chain, will be finitely supported and therefore the distribution function is a piecewise constant function with finitely many jumps. Nevertheless, it is convenient to seek a suitable continuous and continuously increasing interpolation so that its inverse is likewise, in order to avoid some numerical issues. Gaussian kernel estimation was our method of choice justified by its good empirical performance. Thus we have an estimate $\FA \approx F^{-1}_N$ that we calculate ahead of time  as a pre-processing step, after $N>>1$ steps of the gaussian kernel density estimation scheme. (See \cite{Jiang} for a recent work on convergence rate of kernel density estimation.) We use $F^{-1}_N$ as a surrogate for $\FA$.\\

An important difference between ACVaR and the classical CVaR for a single random variable is  that the former involves an arithmetic mean of several random variables and therefore allows a few of them to violate the lower bound in the conditioning, as opposed to a hard constraint in the classical case. Another difference is that while it satisfies translational invariance, positive homogeneity and monotonicity requirements of a coherent risk measure \cite{Artz}, it may not satisfy sub-additivity and therefore convexity in general.\\

The algorithm for ACVaR estimation, adapted from \cite{Kherani}, is a two time-scale stochastic approximation scheme (\cite{BorkarBook}, Chapter 8), with stepsizes $a(n),b(n)$ satisfying 
    \begin{align*}
        &\sum_n a(n) = \infty , \ \sum_n b(n) = \infty, \\
        &\sum_n a(n)^2 < \infty , \ \sum_n b(n)^2 < \infty, \text{ and, } \frac{b(n)}{a(n)} \to 0.
    \end{align*}
    We also have access to a simulated Markov chain $\{X^{orig}_n\}$ on $\mathcal{S}$ with transition probabilities $p(\cdot,\cdot)$. If the reward profile $g(i)$ for each state $i$ is available, then we directly estimate the stationary distribution of the reward. If not, we perform a warm-start by simulating the Markov chain for (say) 10 times the number of states in order to get an estimate of the rewards for each state. If any state is unexplored, we nominally assign it a zero reward. We  simulate another Markov chain $\{X_n\}$ with transition probabilities $p_n(\cdot,\cdot) :=$ estimates for the  tilted probabilities $p^*(\cdot,\cdot)$.  Here $p_n=p(\cdot,\cdot)$ for $n=0$. $p_n(\cdot,\cdot) :=$  the estimated tilted transition probability function for $n \geq 1$.\\
    
    \begin{itemize}
    \item \textbf{Update of tilted transition probabilities:} From the reward $g(k)$ obtained from simulating the chain $\{X_i\}$ as per transition probabilities $p_n(\cdot,\cdot)$ with the last realised state $k$, update the kernel $p_n(\cdot,\cdot)$ according to
    \begin{equation*}
        p_{n+1}(k,l) = \frac{e^{\zeta_n g(k)}}{V_n(i_0)V_n(k)} p(k,l) V_n(l).
    \end{equation*}
    Normalize the row  $p_{n+1}(k,\cdot)$ to ensure that the probabilities sum to $1$. Here $i_0\in \mathcal{S}$ is a fixed state. This is simply the current estimate of the tilted probability $p^*(k,\cdot)$ based on the current estimates $V_n,\zeta_n$ of $V^*, \zeta^*$. \\
    
    \item \textbf{Simulation Step:} If the reward profile is not  provided, simulate a Markov chain $\{X^{orig}_n\}$ with  transition probability $p(k,\cdot)$ and another Markov chain $\{X_n\}$ with tilted probability $p_n(k,\cdot)$ in order to obtain $X^{orig}_{n+1}$ and $X_{n+1}$ respectively. 
 Optional: Continue to update the reward profile CDF by a kernel density estimation scheme and calculate the value of its inverse at the prescribed $c$. If reward profile is provided or estimated as a pre-processing step, then the above can be skipped.  \\  
 
 \item \textbf{Two Time Scale Stochastic Approximation step:}
  Perform two-time scale stochastic approximation for the estimate $V_n$ of $V^*$  and $\zeta_n$ of $\zeta^*$. \\
  
  \textbf{a) The slow time scale:} This is the iteration for $\{\zeta_n\}$ which goes as 
    \begin{equation*}
        \zeta_{n+1} = \Big(\zeta_n + b(n) ( F_N^{-1}(c) - g(X_{n+1}) )\Big)^+,
    \end{equation*}
    where $(\cdots)^+$ on the RHS indicates projection on the positive real line. Note that $(F_N^{-1}(c) - g(X_{n+1}))$ is the empirical gradient of $(\zeta F_N^{-1}(c) - \ln(\rho_{\zeta}))$ as argued in \cite{Kherani}. (See item 5 in the itemized list on p.\ 265 of \cite{Kherani}.) Theoretically, the updates can be performed without projection on the positive real line but practical implementation may encounter instability in initial few iterations due to the exponential presence of $\zeta_n$ in the value function $V_{n+1}$ update. Thus, this is stochastic gradient ascent for evaluating  $\sup_\zeta{(\zeta F^{-1}(c) - \ln(\rho_\zeta))}$ as in \cite{Kherani}. Observe that in the present work, we concern ourselves with positive deviations from the mean ($\forall c>1/2$), at the right tail of reward distribution. For simulating negative deviations on the left tail of the reward distribution ($\forall c<1/2$), $\zeta_n$ should be projected on the negative real line. \\
    
\textbf{b) The fast time scale:} Concurrently, compute
    \begin{equation*}
        V_{n+1}(k) = V_n(k) + a(\nu(k,n))\times\Big[ \frac{e^{\zeta_n g(k)}}{V_n(i_0)} \frac{p(k,X_{n+1})}{p_n(k,X_{n+1})}V_n(X_{n+1}) - V_n(k)\Big].
    \end{equation*}
    Here $\nu(k,n)$ is the count of visits to  state $k$ in $n$ iterations, which serves as the `local clock' at $k$ (\cite{BorkarBook}, Chapter 6). This iteration is an exact analog of the Q-learning scheme for risk-sensitive control from \cite{BorkarQ}  except for the additional importance sampling factor $\frac{p(k,X_{n+1})}{p_n(k,X_{n+1})}$, which is the likelihood ratio that corrects for  using a different transition probability for simulation. The convergence of this scheme can be established as in \textit{ibid.}\\
    
    In fact, the algorithm of \cite{BorkarQ} is for estimating Q-values, whereas here we are estimating a far simpler object, viz., the solution $V^*$ of a multiplicative Poisson equation which is an eigenvalue equation for a non-negative matrix and therefore can be interpreted as the `value' of a risk-sensitive control problem with fixed policy. It is thus a counterpart of value iteration for policy evaluation in risk-sensitive control under fixed stationary policy, more accurately a reinforcement learning counterpart thereof, equivalently a reinforcement learning counterpart of the power method for computing the principal eigenfunction of a non-negative matrix.\\
    
    \item \textbf{Termination:} Terminate when fluctuations in the calculated values of $\zeta_n$  are below a prescribed small threshold in absolute value, over a prescribed number of iterations.
\end{itemize}

\begin{algorithm}
    \caption{Two-Timescale Stochastic Approximation for ACVaR Simulation}
    \label{alg: rare-event-val-iteration}
    \begin{algorithmic}[1]
	 \STATE \textbf{Input}: Markov Chain state space $\mathcal{S}$, transition probabilities $p(\cdot,\cdot)$, stepsize schedules $a(n),b(n)$, reward profile $g(i):\mathcal{S}\to\mathbb{R}$ of the MC (perform initial exploratory simulation to determine $g(\cdot)$ if not already provided).\\
    \textbf{Initialize:} Initialize state reward function $V_0(\cdot)\equiv1$, $p_0(\cdot,\cdot) = p(\cdot,\cdot)$, $\zeta_0 = 1$, $\nu(\cdot,0) = 0$ for all states.
    \STATE Simulate MC for $N>>1$ steps and perform Gaussian KDE to obtain reward distribution. Compute $F_N^{-1}(\cdot)$.
    \FOR {$t = 1, \ldots, T$}
    \STATE Update tilted kernel $p_{t+1}(k,l) = \frac{e^{\zeta_t g(k)}}{V_t(i_0)V_t(k)} p(k,l) V_t(l)$ for current state $k$
    \STATE Obtain new states $X^{orig}_{t+1}$ and $X_{t+1}$ by simulating with transition probabilities $p(k,\cdot)$ and $p_{t+1}(k, \cdot)$
    \STATE $\zeta_{t+1} = \Big(\zeta_t + b(t) ( F_N^{-1}(c) - g(X_{t+1}) )\Big)^+$
    \STATE $V_{t+1}(k) = V_t(k) + a(\nu(k,t)) \Big[ \frac{e^{\zeta_t g(k)}}{V_t(i_0)} \frac{p(k,X_{t+1})}{p_t(k,X_{t+1})}V_t(X_{t+1}) - V_t(k)\Big]$
    \ENDFOR
    \STATE \textbf{Return:} $p_T(\cdot,\cdot)$
\end{algorithmic}
\vspace{-.03in}
\end{algorithm}

\section{Experimental Results}\label{sec:experiments}

Our numerical simulations are for state spaces with $40, 75$ and $150$ states. The original transition probability matrix is randomly generated by adding a small constant value $0.5$ to each $\text{Unif}(0,1)$ entry and normalizing the rows. The reward associated with each state is proportional to its index, scaled appropriately to have the parameter $\zeta_n$ and the average reward in the same order of magnitude while training. We initially perform a “warm-start” or “burn-in” simulation of the Markov chain to explore the reward profile of the state space and get an estimate of Inv-CDF $F^{-1}_N(c)$. The bandwidth parameter of the Gaussian kernel is set to $0.02$, a value smaller than the smallest difference between state rewards, so as to maintain multimodality of the approximation (any other value satisfying this condition also works, with almost indistinguishable results for large state spaces). \\

Following the initial estimation of $F^{-1}_N(c)$, we consider it as having approximately converged to the actual value (by the strong law of large numbers) in $10K$ steps and do not update it later on during the stochastic update iterations (almost identical results are obtained even if we update $F_n^{-1}(c)$ estimate beyond $10K$ steps). We then perform the adaptive update scheme to obtain an estimate of the tilted transition probabilities. The simulation steps are performed with varying step-sizes which are in agreement with the two time-scale stochastic approximation formalism mentioned earlier.  Step sizes are controlled by a single scaling parameter $k$ as
\begin{equation*}
    a(n) = \frac{k}{1+(n+1)^{0.6}}, \quad \quad b(n) = \frac{k}{1+(n+1)^{0.8}}.
\end{equation*}

We observe convergence for all the cases in under $80K$ steps, with (possibly) varying step size parameter $k$ required for different state spaces and different values of $c$. \\

For the sake of brevity, we present here a comparison of $70^{th}$, $85^{th}$, $90^{th}$ percentile thresholding in a $40$ State Markov chain along with the $90^{th}$ percentile thresholding in $75$ and $150$ state Markov chains. Each graph has a value plot on the top depicting the evolution of the parameter $\zeta_n$, the average reward along with $F^{-1}_N(c)$ threshold plotted across number of iterations. The second plot is a frequency diagram which depicts the number of times each state was realized by following the original transition function and the twisted kernel (states have been displayed as per increasing order of reward value). The red vertical line in frequency plots is the $F^{-1}_N(c)$ threshold. \\

Note that the results presented here are for rewards defined in an arithmetic progression. We also tested our method on deterministic rewards that are chosen from a $\text{Unif}(0,4)$ distribution initially, obtaining similar convergence results. The scaling of reward profile to maintain $\zeta_n$ values and average reward of the same order of magnitude reduces the possibility of numerical instability. Although theoretical guarantees are provided for all finite-valued reward profiles, we restrict ourselves to the above in order avoid the additional burden of fine-tuning the step-size schedules. \\

\begin{figure}
   \centering
   \includegraphics[height=3.3in]{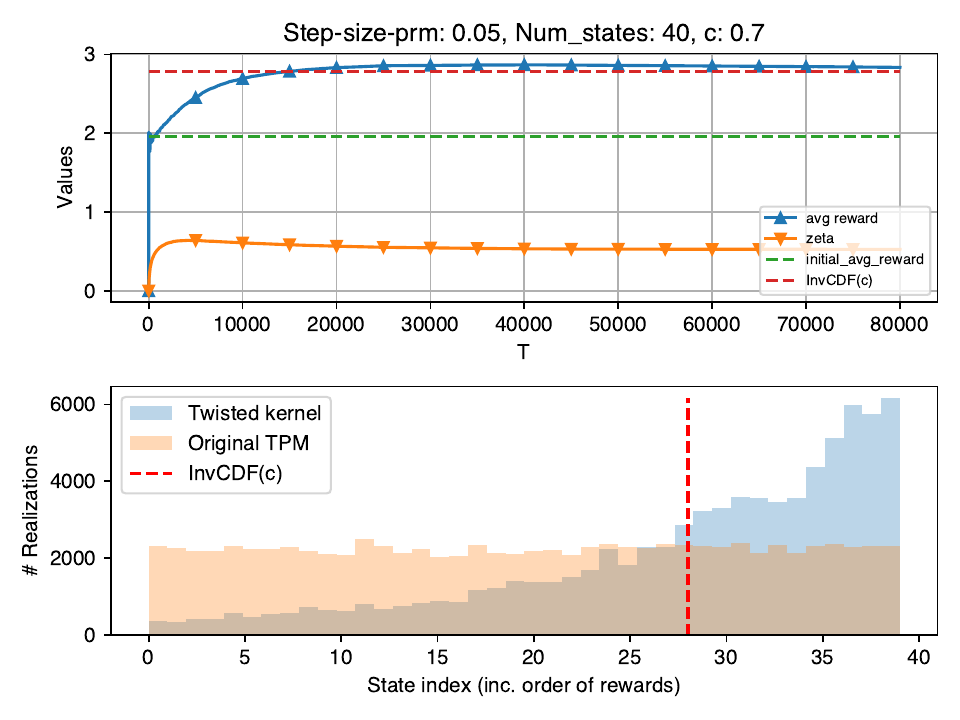}
   \caption{40 State MC conditioned above $70^{th}$ percentile average outcomes}
\end{figure}
\begin{figure}
   \centering
   \includegraphics[height=3.3in]{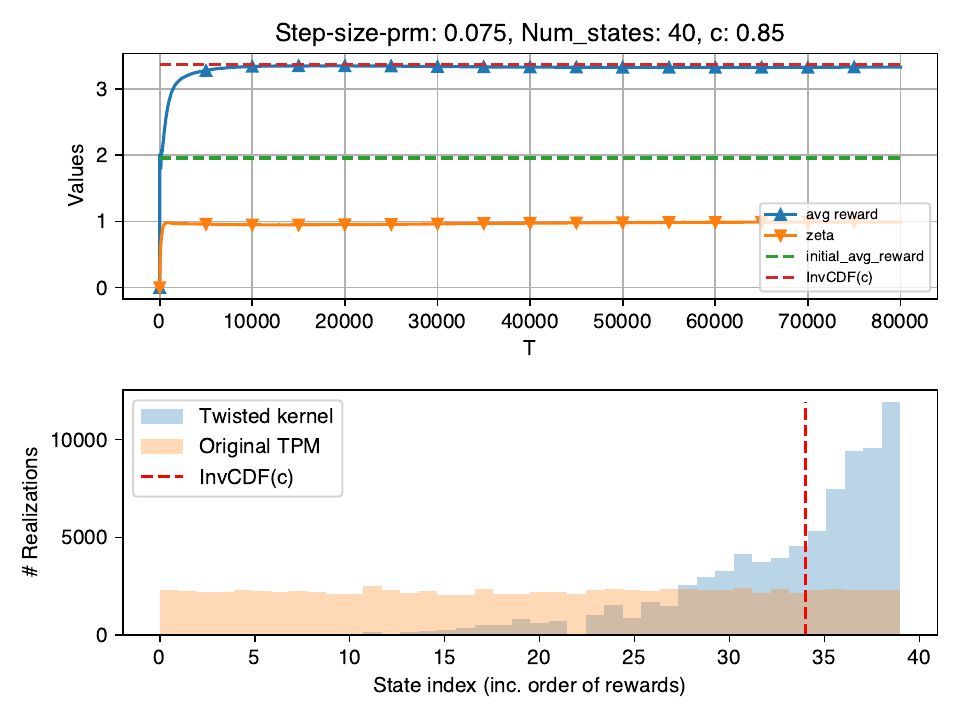}
   \caption{40 State MC conditioned above $85^{th}$ percentile average outcomes}
\end{figure}

\begin{figure}
   \centering
   \includegraphics[height=3.3in]{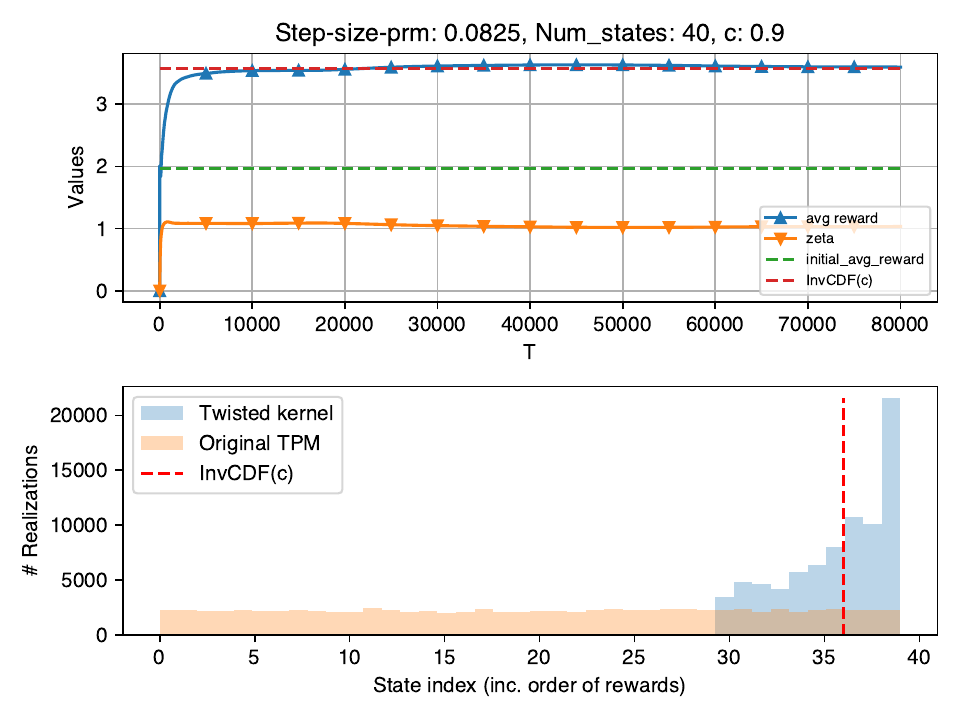}
   \caption{40 State MC conditioned above $90^{th}$ percentile average outcomes}
\end{figure}

\begin{figure}
   \centering
   \includegraphics[height=3.3in]{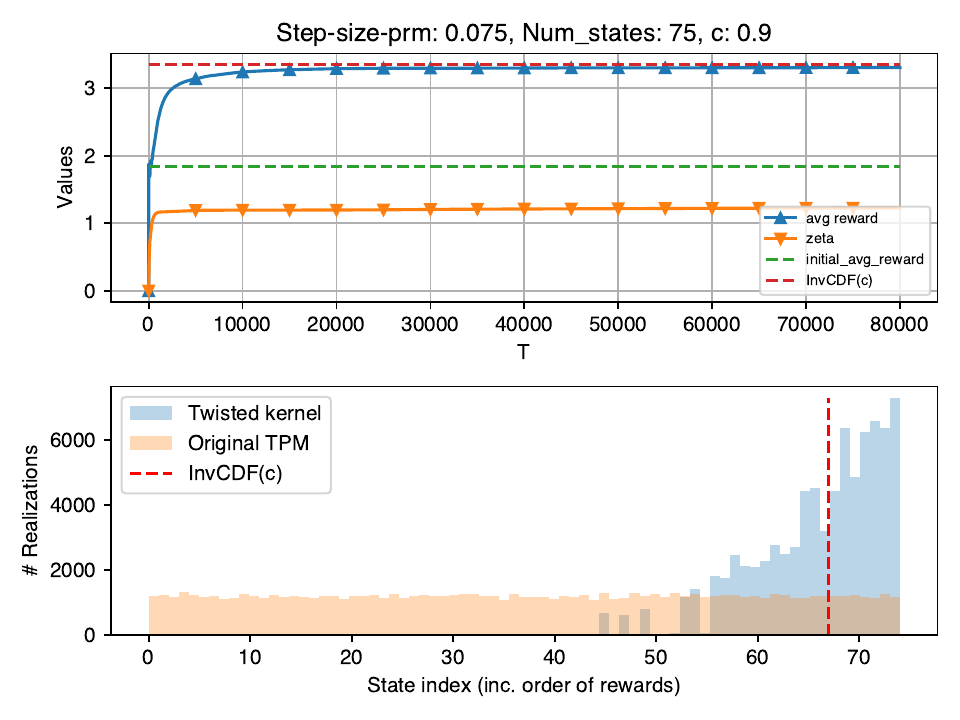}
   \caption{75 State MC conditioned above $90^{th}$ percentile average outcomes}
\end{figure}

\begin{figure}
   \centering
   \includegraphics[height=3.3in]{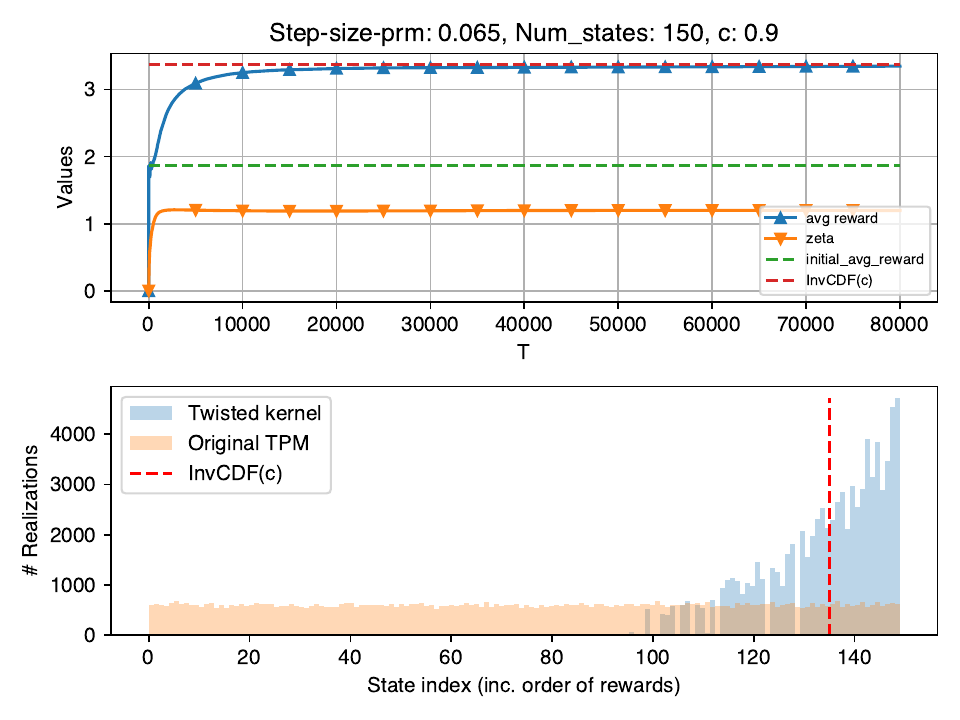}
   \caption{150 State MC conditioned above $90^{th}$ percentile average outcomes}
\end{figure}

\section{Conclusion}\label{sec:conclusion}

We have proposed an asymptotic variant of the classical CVaR, dubbed ACVaR, and a stochastic approximation scheme for its computation. It is motivated by its apparent similarity with CVaR, relative ease of computability, and theoretical justification derived from the developments of \cite{Kherani}. This is a first step in this direction and is far from being a closed topic. Some technical issues that remain are as follows.

\begin{enumerate}
\item Can the double limit in \eqref{acvar} be replaced by a single limit by setting $m = n$?
\item Can $F^{-1}(c)$ in \eqref{acvar} be replaced by $F_n^{-1}(c)$, this term thus becoming a part of the limiting process?
\item Can one modify the definition in such a way that all the axioms of \cite{Artz} for a coherent risk measure hold?
\item Is there an estimation scheme for estimating inverse CDF evaluated at $c$ with $\mathcal{O}\left(\frac{1}{n}\right)$ decay so that the estimation can be made a concurrent part of the algorithm after the warm start, in a manner that allows us to leverage the results of \cite{Kherani}, \cite{Konto} to include it in the overall analysis?
\end{enumerate}

\section*{Acknowledgement}
VB was supported in part by a grant from Google Research, Asia.

\end{document}